\documentclass{iau} 
\usepackage{graphicx}

\title[Pulsar glitch and nuclear EoS] 
{Pulsar glitch and nuclear EoS:\\Applicability of superfluid model}

\author[Ang Li \& Rui Wang]   
{Ang Li \and Rui Wang}
 
\affiliation{Department of Astronomy, Xiamen University, Xiamen, Fujian 361005, China \\ email: {\tt liang@xmu.edu.cn} }

\pubyear{2017}
\volume{337}  
\jname{Pulsar Astrophysics -­ The Next 50 Years}
\editors{P. Weltevrede, B.B.P. Perera, L. Levin Preston \& S. Sanidas, eds.}

\begin{document}

\maketitle

\begin{abstract}
In this contribution, we have made use of the contemporary equation of states (EOSs) for the complete neutron star structure, and confronted them with one particular glitch constrain for the crustal moment of inertia (MOI). We find that with these EOSs, the radii of three millisecond pulsars selected by NICER: PSR J0437-4715, PSR J1614-2230, PSR J0751+1807, are all around 12.5 km. Also, a star with $ M \lesssim 1.55 M_{\odot}$ would fulfill the MOI calculation for glitch constrain with the latest neutron superfluidity density, and the glitch crisis might not be present.

\end{abstract}

              

Glitch, the sudden jump of pulsar rotational frequency, has become one of the most important observables that helps us unravel the unknown equation of state (EOS) of neutron stars (NSs). It is generally believed to be related to the dynamics of the internal rotating components, and as evidence of superfuid components. This is the widely-known two-component mechanism or superfluid model. The latest ``glitch crisis'' problem in the standard superfluid model has generated a great debate. The conclusion is still puzzling. The present contribution will focus on some recent progresses on the crisis study, especially with the most updated unified NS EOSs and neutron superfluid density. 


We select unified NS EOSs that satisfy up-to-date experimental constraints from both nuclear physics and astrophysics, namely BCPM (\cite[Sharma \etal\ 2015]{Sharma_etal15}), BSk20, BSk21~(\cite[Potekhin \etal\ 2013]{Potekhin_etal13}). The possible effects of pasta phase~ (e.g., \cite[Hooker \etal\ 2015]{Hooker_etal15}) and strangeness phase transitions in the stars' core (e.g., \cite[Li \etal\ 2015]{Li_etal15}; \cite[Zhu \etal\ 2016]{Zhu_etal16}) are referred to a future study. 

The resulting mass-radius relations with the employed EOSs are shown in Figure 1. Adapting from the left panel of Figure 2 of Li et al., (2016a), we add three shaded areas indicating the measured masses of three millisecond pulsars selected by NICER: PSR J0437-4715, PSR J1614-2230, PSR J0751+1807. Their rotational frequencies are 173.6 Hz, 317.5 Hz, 287.4 Hz, respectively, much smaller than the corresponding Kepler frequency predictions [above 1 kHz (e.g., \cite[Li \etal\ 2016b]{Li_etal16b})]. Therefore the slow-rotation approximation can be applied and we expect only small increase of the gravitational mass/radius by rotation for these stars. Thus the calculated results from the static TOV equation in Figure 1 can be regarded as fairly good estimations for the stars' radii. We mention here that their radii are all around 12.5 km based on one of the unified EOSs, BSk21, whose high-density EOS part reproduces the microscopic calculations.

\begin{figure}[h]
\begin{minipage}{15pc}
\includegraphics[width=16pc]{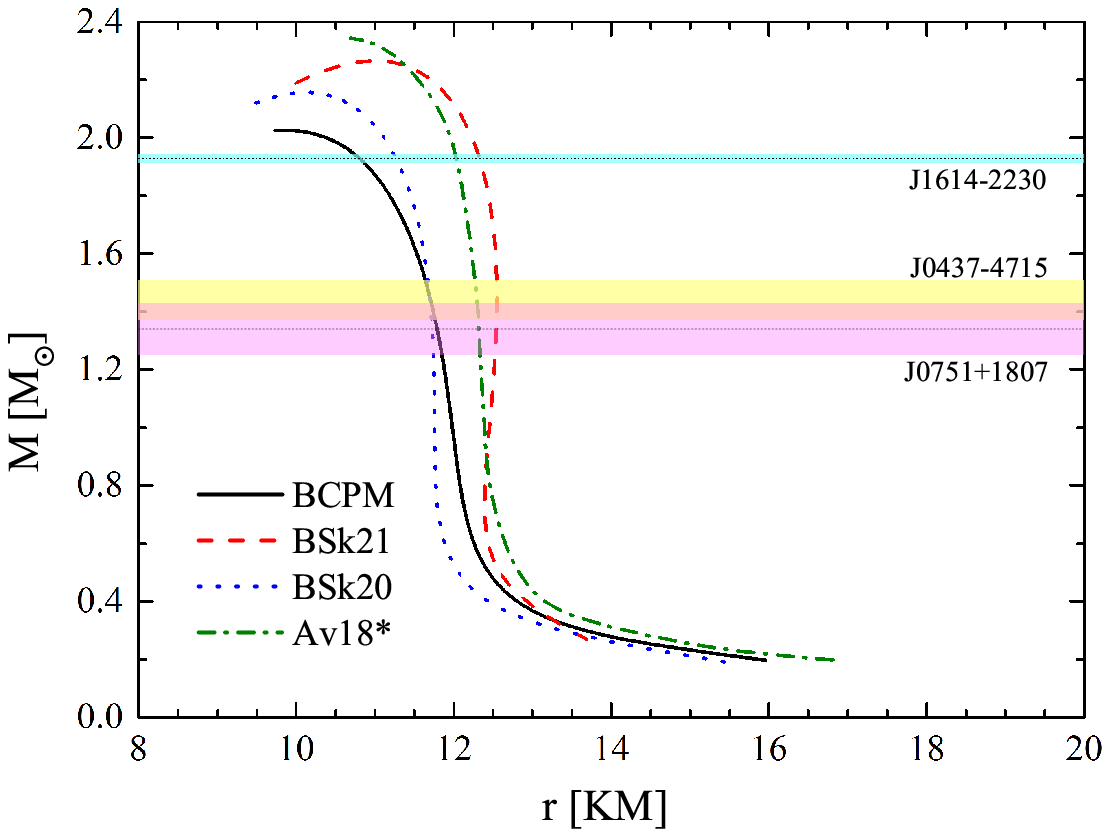}
\caption{Mass-radius relations of the star, for four cases of NS EOSs (BCPM, BSk21, BSk20, Av18*). ``Av18*'' stands for the matched ``BHF + NV + BPS'' EOS. The shaded areas are for three millisecond pulsars selected by NICER with known masses, respectively. Adapted from Li et al., (2016a).}
   \label{fig1}
    \end{minipage}\hspace{1pc}%
\begin{minipage}{15pc}
 \includegraphics[width=16pc]{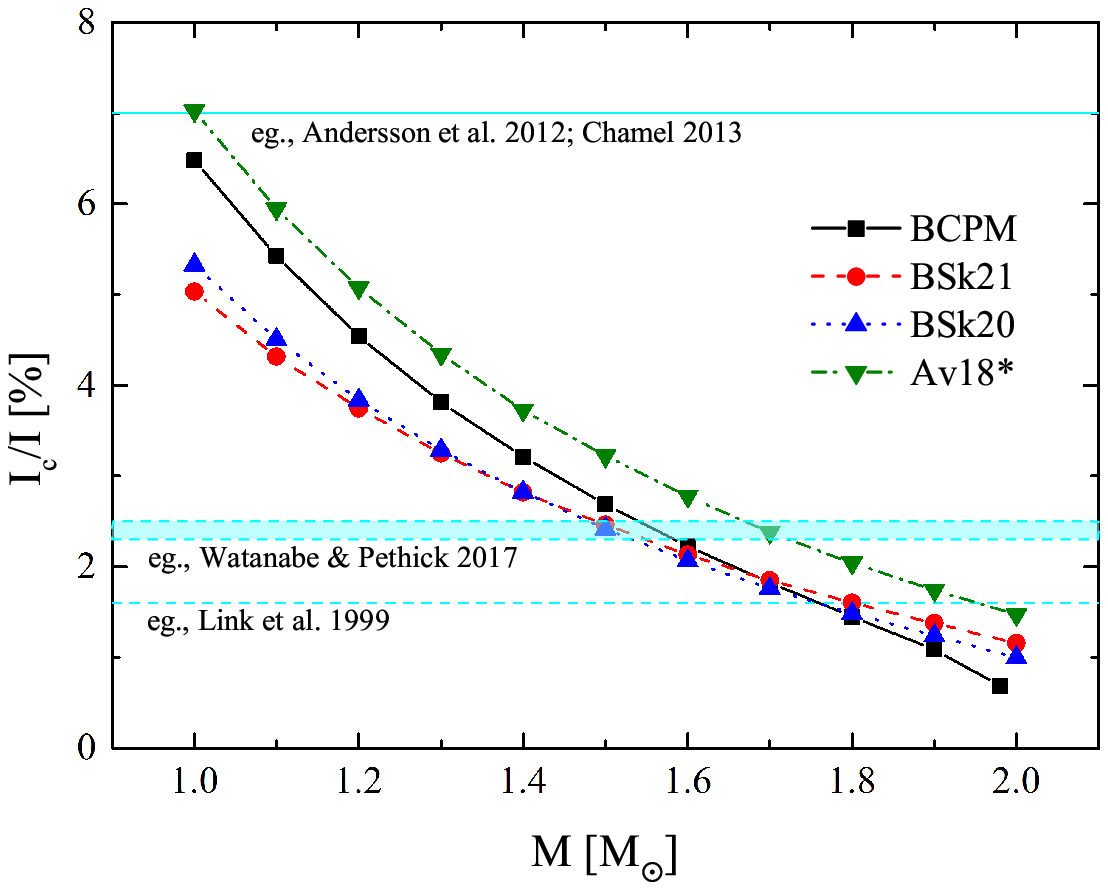} 
 \caption{Fractional moments of inertia as a function of the stellar mass for the same EOS group with Figure 1. Three horizontal lines are three glitch constraints from the labelled articles, respectively. Adapted from Li et al., (2016a).}
   \label{fig2}
    \end{minipage}
\end{figure}

The “glitch crisis” problem is related to the moment of inertia (MOI) of the stellar crust, e.g., \cite[Li (2015)]{Li15}; \cite[Li \etal\ (2016a)]{Li_etal16a}. The glitch observations from the Vela pulsar has put a constrain on the fractional MOI, which is $I_c/I \ge 1.6\%$ (e.g., \cite[Link \etal\ 1999]{Link_etal99}). The crisis refers to the lift of the lower limit from $1.6\%$ to $\sim 7\%$ (e.g., \cite[Andersson \etal\ 2012]{Andersson_etal12}; \cite[Chamel \etal\ 2013]{Chamel_etal13}), which makes it very difficult for the nuclear EOSs to fulfill with a normal $M > 1.0 M_{\odot}$ NS. This is clearly shown in Figure 2. However, a recent calculation (\cite[Watanabe \& Pethick 2017]{Watanabe17}) has brought the problem of the crustal neutron superfluid density to our attention. They argued that the previous value, because of the omit of pairing, might overestimate the entrainment effect. With a typical neutron gap of $\Delta = 1 - 1.5$ MeV, they obtained the conduction neutrons would be $64\% - 71\%$ of the total unbound neutrons, not as large as the previous $23.3\%$ (\cite[Chamel \etal\ 2013]{Chamel_etal13}). Those result in a new constrain for the fractional MOI of $I_c/I \ge 2.5 - 2.3 \%$. Then a NS of $M \lesssim 1.55 M_{\odot}$ NS would satisfy our employed EOSs. Most of the NSs with known masses have masses in this range (e.g., \cite[Lattimer 2012]{Lattimer12}). However, more detailed work has to be done to solve the crisis problem (see discussions in e.g., \cite[Chamel 2017]{Chamel17}).

The contribution was supported by the National Natural Science Foundation of China (No. U1431107), and the Fundamental Research Funds for the Central Universities.

\end{document}